\documentclass[aps,prl,twocolumn,superscriptaddress]{revtex4-1}

\usepackage{amsfonts}
\usepackage{amsmath}
\usepackage{amssymb}
\usepackage{graphicx}
\usepackage{float}
\usepackage[dvipsnames]{xcolor}
\begin{document}
\title{Dynamic phase transition induced by active molecules simulating a facilitation mechanism in a supercooled liquid} 

\date{\today }

\author{Victor Teboul}
\email{victor.teboul@univ-angers.fr}
\affiliation{Laboratoire de Photonique d'Angers EA 4464, Universit\' e d'Angers, Physics Department,  2 Bd Lavoisier, 49045 Angers, France}


\begin{abstract}
%Facilitation mechanisms, that is the fact that mobile molecules facilitate the motion around them, while non mobile ones hinder it, are supposed to be at the origin of the appearance of dynamic heterogeneity in supercooled liquids in their approach to the glass transition.
%(nucleation similar crystallisation)
 The purpose of this work is to use active particles to study the effect of facilitation on supercooled liquids.
To this end we investigate the behavior of a model supercooled liquid doped with  intermittently active and intermittently slowed particles.
To simulate a facilitation mechanism, the active particles are submitted intermittently to a force following the mobility of the most mobile molecule around, while the slowed particles are submitted to a friction force.
%We study the modification of the medium and active particles properties when the characteristic time $\tau_{\mu}$ used for the definition of the mobility increases.
We observe upon activation, a fluidization of the whole medium simultaneously to a large increase of the dynamic heterogeneity.
This effect is reminiscent of the fluidization observed with molecular motors doping of the medium. %, with the induced dynamic heterogeneity characteristic time being controlled by the activation mechanism.
%We observe an acceleration of these mechanisms when $\tau_{\mu}$ is increased, and a decrease of the characteristic time scales of the material.
When the mobility characteristic time $\tau_{\mu}$ used in the facilitation mechanism matches the physical time $t^{*}$ characterizing the spontaneous mobility aggregation of the material, we observe a phase transition associated with a structural aggregation of active molecules.
This transition is accompanied by a sharp increase of the fluidization and dynamic heterogeneity.

\end{abstract}

\maketitle
\section{ Introduction}

Active matter, that is matter doped or constituted of particles that are able to move by themselves, like molecular motors or nano-machines\cite{motoro1,motoro2,motoro3,motoro4,motoro5,motoro6,motoro7,motoro8,motoro9,motoro10,motoro11,motoro12,motoro13,motoro14,motoro15,motoro16,motoro17,prefold,us1}, is retaining much attention by the scientific community due to its connection with biology and out of equilibrium statistical physics\cite{active1}.
{\color{black} Active matter appears also as a new route to study the glass transition problem\cite{gt0,gt1,gt2,anderson,fragile1,ms1,ms2,ms3,ms4,ms5}, due to its non equilibrium physical characteristics and possible origin.}
That direction of research already lead to a number of interesting results\cite{active1,active2,active3,active4,active5,active6,active7,active8,active9,active10,active11,active12,active13,active14,active15,active16,active17,active18,active19,Szamel1,Szamel2,Szamel3,Szamel4,Szamel5,Szamel6,Szamel7}.
It was found that the glass transition persists in active matter although at a different glass transition temperature.
In specific conditions a dynamical slowing down appears upon activation, while in most conditions a fluidization is observed\cite{Szamel1,Szamel2,Szamel3,Szamel4,Szamel5,Szamel6,Szamel7}.   

An induced fluidization was also reported\cite{flu1,flu2,flu3,flu4,flu5,flu6,md16,cage,prefold,rate,us1} experimentally and with simulation when simple molecular motors\cite{az1,az2,az3,az4,az5,az6,az7,az8,az9,az10} activate a soft material.
 The relation between these two fluidization mechanisms is however not established.
The fluidization induced by molecular motors is due to the activation of the spontaneous (thermal) cooperative mechanisms in supercooled mediums.
Molecules pushed away by the nano-motor induce molecular motions around them due to the medium's cooperativity. 

Here we raise the question of inducing new cooperative mechanisms inside the medium using active molecules moved by forces which directions depend on the mobility of their surroundings.
We expect the facilitation created in this way to induce cooperativity, resulting in a fluidization of the medium.

An usual way to create active media in simulation is to add a propulsive force on molecules most often in the direction of their velocities.
Sometimes an interaction pushing to the alignment of molecules displacements is added, to create complexes behaviors resembling biological displacements.
In this work we study {\color{black} the effect of forces which characteristics better correspond to the physics of supercooled liquids.}
Therefore our propulsive force will follow the mobility of molecules instead of their velocity, and correlation between velocities will be replaced by correlation in mobilities of molecules. To summarize, we expect that activating our medium with relevant parameters, will increase (or decrease) the thermal cooperativity of the medium in its approach to the glass transition, resulting in a modification of the transition and physical properties of the medium, in particular a fluidization, that we will study.
%Another question of interest is the effect of activation on medium's crystallization mechanism.

\section{Calculation}

{\color{black} Our purpose in this study is to test an out of equilibrium model mimicking the behavior of supercooled liquids.}
We use out of equilibrium molecular dynamic simulations\cite{md1,md2,md2b,Aurelien} in order to find results easier to understand in a practical way and easier to compare to previous studies on molecular motors\cite{md16,rate,c3,c4,pccp,carry}, in particular to fluidization phenomena observed experimentally and theoretically\cite{md16,flu1,flu2,flu3,flu4,flu5,cage,flu6} when molecular motors are diluted inside soft materials.
%In addition to theoretical and experimental methods, simulations in particular  molecular dynamics and Monte Carlo simulations\cite{md1,md2,md2b,md4} together with model systems\cite{ms1,ms2,ms3,ms4,ms5}  are now widely used to unravel unsolved problems in condensed matter and complex systems physics\cite{keys,md3,md4b,md6,md7,md8,md9,ee2,md11,md12,md13,md14,md15,ee1,finite1,u1,Aurelien}. 
The system is maintained out of equilibrium by the presence of active molecules releasing energy into it and damped molecules absorbing energy.
As a result our system with an adequate choice of the damping parameter while out of equilibrium is approximately in a steady state, the damped molecules removing the energy given to our system by the activated molecules. We add anyway a small thermostat\cite{berendsen} to avoid any possible drift in energy. 
In our calculations, $10$ percents of the medium molecules are damped during $10 ps$ taken at random but continuously in a $40 ps$ time lapse.
Simultaneously $10$ percents of the medium molecules are activated (pushed) in the direction of the most mobile of their neighbors during $10 ps$ also taken at random but continuously in a $40 ps$ time lapse. Therefore, at any time $2.5$ percent of molecules are damped and $2.5$ percents are accelerated, while the other $95$ percent molecules do not experience external forces but only the intermolecular interactions. 
We define the mobility $\boldsymbol \mu_{i}(t)$ of a molecule $i$ as 
\begin{equation}
\displaystyle {\boldsymbol \mu_{i}(t)={\bf r}_{i}(t+\tau_{\mu})-{\bf r}_{i}(t)}    \label{e0}
\end{equation}
where $\tau_{\mu}$ will be called  the mobility characteristic time.
By tuning $\tau_{\mu}$ we expect to find a significant response of our liquid when $\tau_{\mu}$ will match an important characteristic time scale of the physics of our medium.
Activated molecules are periodically subject to a force ${\bf {f}}_{i}^{a}$ of constant intensity $f_{0}$, acting during $10 ps$ of a time period $T=40ps$. The forces ${\bf{f}}_{i}^{a}$ are limited to a set of molecules called active, and begin with a different random time origin for each molecule, following the law:
\begin{equation}
 {\bf {f}}_{i}^{a}({\tau_{\mu}},t) = {{ f_{0} {\theta}_{i,T,{\Delta T}}(t)  {\bf{u}}_ {i,{\mu}_{max}}^{neighbor}}(t,{\tau}_{\mu})}            \label{e1}
\end{equation}
Here ${\theta}_{i,T,{\Delta T}}(t)$ is a periodic heaviside function, equal to $1$ during $\Delta T=10ps$ and zero otherwise with a period $T=40ps$.
$f_{0}=6.02$ $10^{-14} N$ is the constant intensity of the force when activated.  
\begin{equation}
{ {\bf{u}}_ {i,{\mu}_{max}}^{neighbor}(t,{\tau}_{\mu})=\boldsymbol \mu_{j}(t,{\tau}_{\mu})/\lvert \mu_{j}(t,{\tau}_{\mu})\rvert}        \label{e1b}
\end{equation}
 is the unit vector of the mobility of the most mobile neighbor $j$ of molecule $i$.

\vskip0.5cm
The damped molecules are subject to a force ${\bf{f}}_{i}^{d}$ proportional to their velocity:
\begin{equation}
 {\bf {f}}_{i}^{d}(t) = {{ - \alpha.f_{0} {\theta}_{i,T,{\Delta T}}(t)  {\bf{v}}_ {i}}(t)/{\bar{ v  }}}            \label{e2}
\end{equation}
Where $\bar{v }=\sqrt{8 RT \over \pi m}$ is the average velocity modulus, $m=80g/mol$ is the molar mass of a molecule and the temperature is $T=500K$.
$\alpha= 9.1$ is a coefficient chosen so that the energy absorbed by damped molecules approximately compensate in our system the energy released by active molecules.
Our medium is a minimal model liquid\cite{ariane} constituted of dumb bell diatomic molecules (each atom being of the same mass $m_{0}=40g/N_{A}$) chosen to hinder crystallization\cite{mix1,mix2} and accelerate the simulations.  However due to the use of Lennard Jones potentials only, the results can be easily shifted to model although approximately a large number of real viscous liquids.
  In this work we study the modification of the main characteristics of supercooled liquids, the presence of dynamic heterogeneity, the diffusion properties, and the $\alpha$ relaxation time related to the viscosity of the medium. 
We will now define the statistical functions used in that purpose.
The most adequate function to measure the strength of the dynamic heterogeneity\cite{dh0,dh1,dh2,dh3} is the dynamic susceptibility $\chi_{4}$ defined as\cite{dh0}:

\begin{equation}
\chi _{4}(a,t)=\frac{\beta V}{N^{2}}\left( \left\langle
C_{a}(t)^{2}\right\rangle -\left\langle C_{a}(t)\right\rangle ^{2}\right) 
\label{e2}
\end{equation}
with 
\begin{equation}
C_{a}(t)={\sum_{i=1}^{N}{w_{a}}}\left( \left\vert {{{\mathbf{r}}}}_{i}(t)-{{{\mathbf{r}}}}_{i}(0)\right\vert \right) .  \label{e3}
\end{equation}

In these equations, $V$ denotes the volume of the simulation box, $N$ denotes the number of molecules in the box, and $\beta =(k_{B}T)^{-1}$. Also, the symbol $w_{a}$ stands for a discrete mobility window function, $w_{a}(r)$, taking the values $w_{a}(r)=1$ for $r<a$ and zero otherwise. We use the value $a_{0}=1$\AA\ below the transition\cite{transition} or for the non-activated liquid and $a_{1}=2$\AA\ above the transition, values that in these conditions maximize $\chi_{4}(a,t)$.
The Non Gaussian parameter (NGP) $\alpha_{2}(t)$ is also often used as a measure of dynamic heterogeneity.
\begin{equation}
\displaystyle{ {\alpha}_{2}(t)={d\over{d+2}} {<r^{4}(t)>\over{<r^{2}(t)>^{2} }}   -1    }\label{e5}
\end{equation}
Where $d=3$ is the system dimension.
It has the advantage of being simple to interpret and to have a characteristic time $t^{*}$ directly related to the physics of the liquid.
We define $t^{*}$ as the time for which $\alpha_{2}(t)$ reaches its maximum.
Due to the definition of $\alpha_{2}(t)$, in most supercooled liquids $t^{*}$ is the characteristic time of cooperative motions (DHs).
Another function of large interest in glass-transition related phenomena is the intermediate scattering function  $F_{S}(Q,t)$  that represents the autocorrelation of the density fluctuations at the wave vector Q.
This function gives information on the structural relaxation of the material.
We define $F_{S}(Q,t)$ by the relation:
\begin{equation}
\displaystyle{F_{S}(Q,t)={1\over N N_{t_{0}}} Re( \sum_{i,t_{0}} e^{i{\bf Q.(r_{i}(t+t_{0})-r_{i}(t_{0}))}}  )          }\label{e1}
\end{equation}
For physical reasons, Q is chosen as the wave vector (here $Q_{0}=2.25$\AA$^{-1}$) corresponding to the maximum of the structure factor $S(Q)$.
$F_{S}(Q_{0},t)$ then allows us to calculate the $\alpha$ relaxation time $\tau_{\alpha}$  of the medium from the equation: 
\begin{equation}
\displaystyle{F_{S}(Q_{0},\tau_{\alpha})=e^{-1}}       \label{e10}     
\end{equation}
Finally, the diffusion coefficient $D$ is obtained from the long time limit of the mean square displacement $<r^{2}(t)>$:
\begin{equation}
\displaystyle{<r^{2}(t)>=   {1\over N N_{t_{0}}}  \sum_{i,t_{0}}  ({\bf r}_{i}(t+t_{0})-{\bf r}_{i}(t_{0}))^{2}                                }       \label{e11}     
\end{equation}
and
\begin{equation}
\displaystyle{\lim_{t \to \infty}  <r^{2}(t)>=  2 d D t                               }       \label{e12}     
\end{equation}

\section{Results and discussion}

\subsection{Medium fluidization\\}

\begin{figure}%[H]
\centering
\includegraphics[height=6.2 cm]{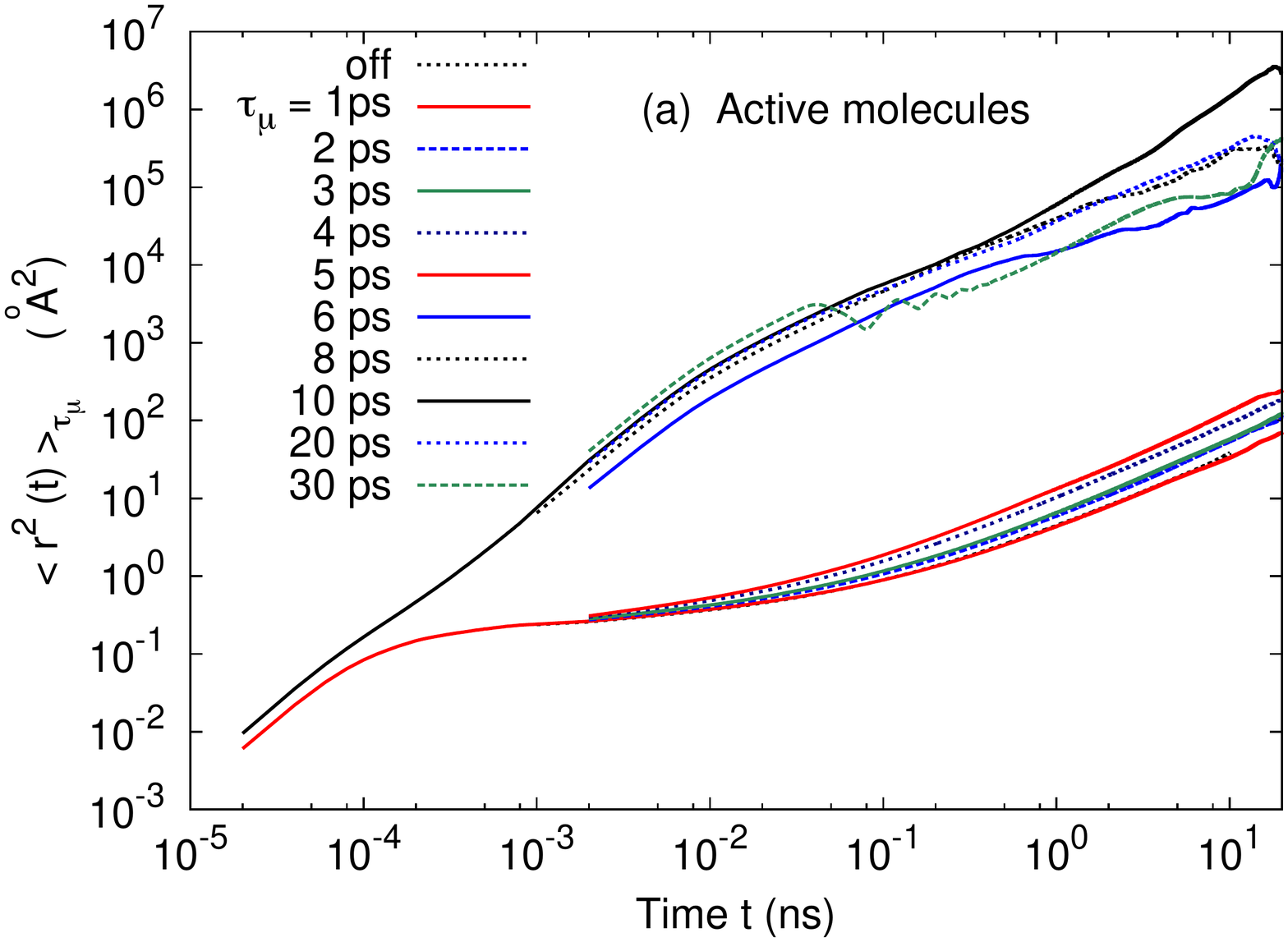}
\includegraphics[height=6.2 cm]{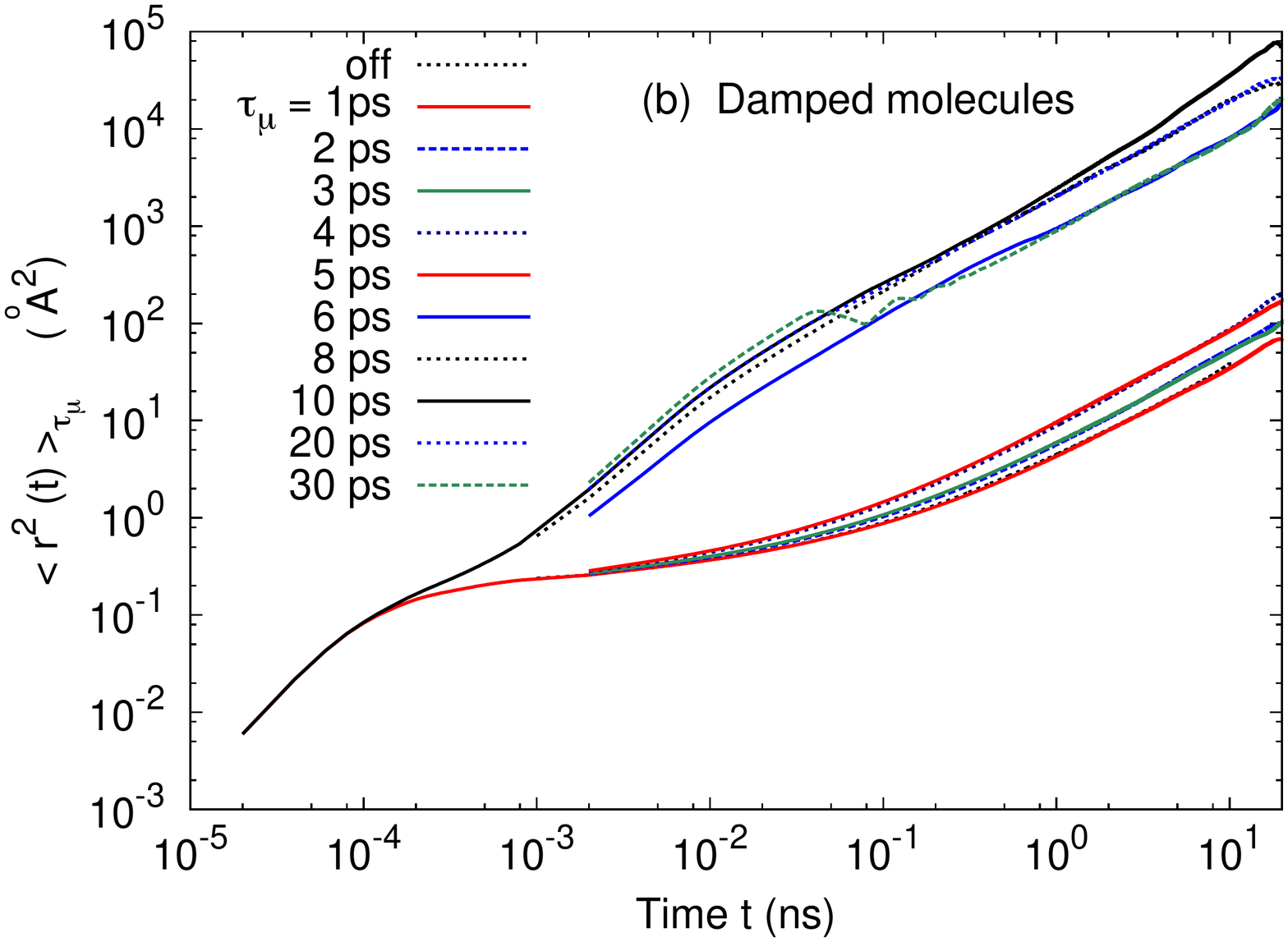}
\includegraphics[height=6.2 cm]{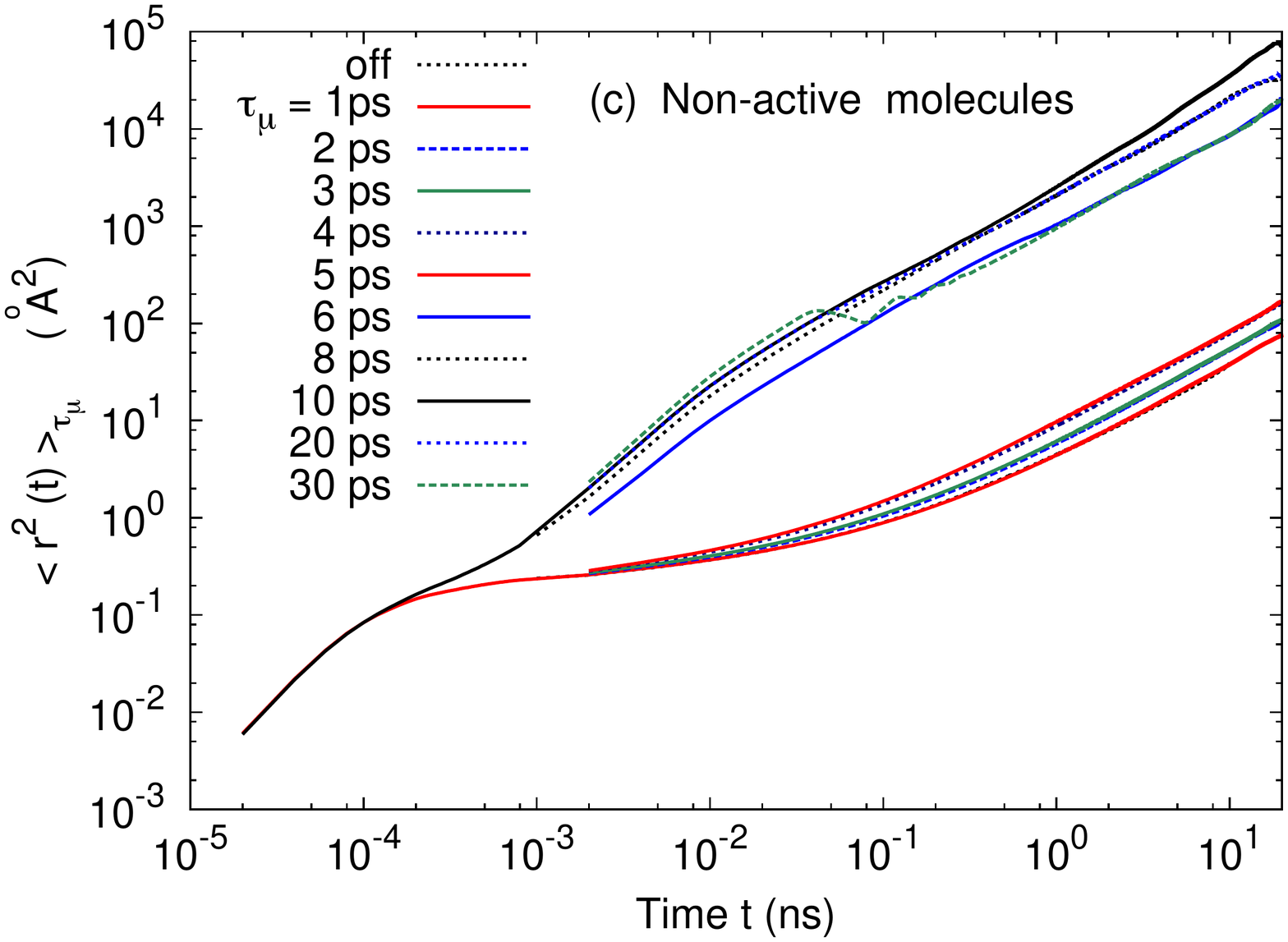}

\caption{{\color{black}(color online)  Mean square displacement $<r^{2}(t)>$ of (a) the $10$ percent intermittently  active molecules, (b) the $10$ percent intermittently damped molecules, (c) the $80$ percent medium molecules moving freely. $\tau_{\mu}$ is the characteristic time chosen for the mobility definition acting in the activation force (see the calculation section for details). Active molecules are activated during $1/4$ of a time period $T=40 ps$ that begins randomly for each active molecule. Consequently there are only an average of $2.5$ percent of molecules active at a given time $t$ and similarly $2.5$ percent of damped molecules.} }
\label{f1a}
\end{figure}

\begin{figure}%[H]
\centering
\includegraphics[height=6.8 cm]{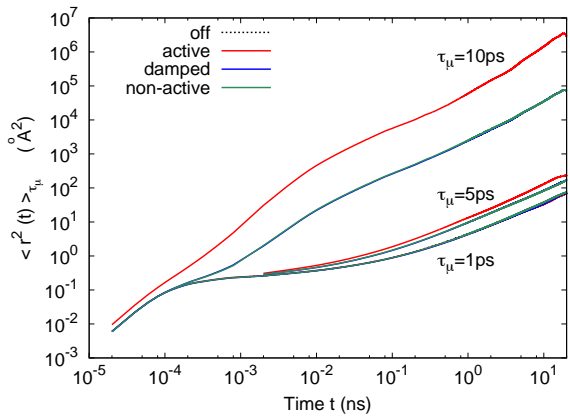}

\caption{{\color{black}(color online)  Mean square displacement $<r^{2}(t)>$ for various values of $\tau_{\mu}$  (the characteristic time chosen for the mobility definition acting in the activation force). 
There is not much difference between active and damped molecules below the transition, while above the transition active molecules are more mobiles due to their aggregation. 
} }
\label{f1d}
\end{figure}

\begin{figure}%[H]
\centering
\includegraphics[height=6.8 cm]{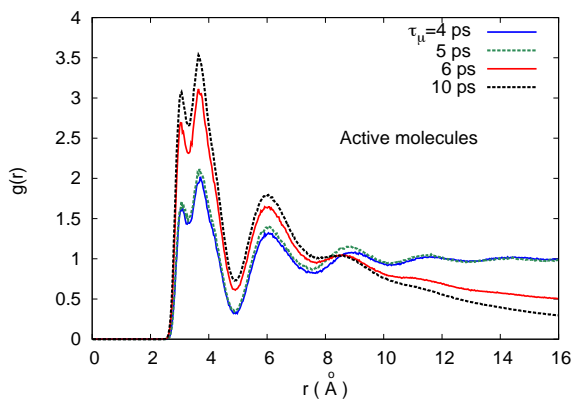}

\caption{{\color{black}(color online)  Radial distribution function $g(r)_{active-active}$ between the $10$ percent intermittently active molecules for different values of the parameter $\tau_{\mu}$.
We observe the aggregation of the active molecules at the phase transition, that is when $\tau_{\mu} \geq \tau_{\mu}^{c} = 5.6 ps$. } }
\label{frdf}
\end{figure}

When active and damped molecules are together introduced inside the medium, we observe its fluidization.
This fluidization increases with the choice of mobility's characteristic time ${\tau_{\mu}}$. 
Figures \ref{f1a} and \ref{f1d} illustrate that behavior displaying the evolution of the mean square displacements $<r^{2}(t)>_{\tau_{\mu}}$ with ${\tau_{\mu}}$, of active, inactives and damped molecules.
A sharp increase appears at the critical value ${\tau_{\mu}}^{c}=5.6 ps$\cite{transition}. 
For the same value of ${\tau_{\mu}}\geq {\tau_{\mu}}^{c}$ we observe a sharp structural aggregation of mobile molecules (see Figure \ref{frdf}) %suggesting
showing a dynamic phase transition associated with the mobility\cite{transition}.
We thus explain the sharp increase of the displacements as due to  the structural aggregation of active molecules that facilitates their motion.

%Why does the characteristic time ${\tau_{\mu}}$ trigger the aggregation of active molecules ? 

Figure \ref{f1a} shows that non only the active molecules undergo a transition in mobility, but also the large set of non-active molecules and even the damped molecules.
In Figure \ref{f1d} we compare the MSD of the different set of molecules (actives, inactives and damped)  in Figure \ref{f1d} for different values of the characteristic time ${\tau_{\mu}}$.
For short values of the characteristic time (${\tau_{\mu}}=1ps$ on the plot), the activation has no effect on the MSDs, then as the characteristic time increases, the active molecules MSD split slightly from the two other curves.
Eventually, when the characteristic time reaches the transition value, the active molecules curve separates itself importantly from the two other curves and we observe a factor larger than 10 for  ${\tau_{\mu}}=10ps$,  while the damped and non-active sets of particles lead unexpectedly to the exact same MSDs.

\subsection{Aggregation of active molecules\\}

The structure of the different sets of molecules (i.e. active, damped and normal) is similar below the transition\cite{transition} i.e. for  ${\tau_{\mu}}\leq {\tau_{\mu}}^{c}  $ to the structure without activation. 
Then at the transition and above the transition, the active molecules aggregate as shown in Figure \ref{frdf}, while the other sets of molecules do not (not shown). 
We interpret this aggregation of the active molecules as the main cause of the acceleration of the liquid dynamics.
 In that picture, because the active molecules are the most mobiles, their aggregation facilitates their motion due to the decrease of their surrounding viscosity.
 
 The characteristics of an aggregation are observed in Figure \ref{frdf} for  ${\tau_{\mu}}= {\tau_{\mu}}^{c}$ and $10 ps$ as the first two peaks that represent the density probability to have active molecules as first and second neighbors increase while the density probability at larger distances decrease.
In contrast, below the transition we observe a radial distribution function very typical for a liquid and that is identical to the RDF of the liquid without activated molecules.
Below the transition, for $r>10$\AA\ the radial distribution  $g(r)\approx1$ showing the homogeneity of the medium after the third neighbor correlation, while the depletion above the transition shows that the active molecules have migrated to shorter distances.

\subsection{Characteristic times evolution\\}

\begin{figure}%[H]
\centering
\includegraphics[height=6.8 cm]{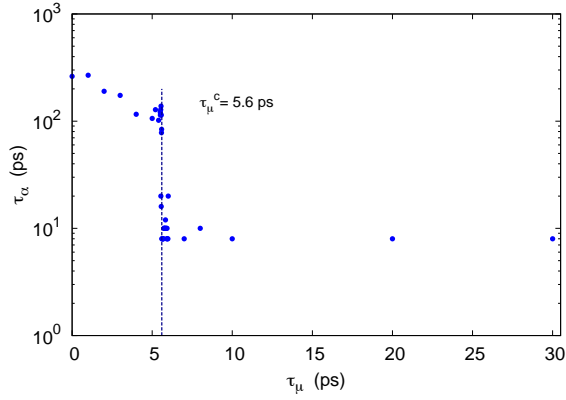}
\caption{{\color{black}(color online)  Alpha relaxation time $\tau_{\alpha}$ as a function of $\tau_{\mu}$. $\tau_{\alpha}$ is here obtained from the relation 
$F_{s}(Q_{0},\tau_{\alpha})=e^{-1}$ where $Q_{0}=2.25$ \AA$^{-1}$.
}}
\label{ftaualpha}
\end{figure}

\begin{figure}%[H]
\centering
\includegraphics[height=6.8 cm]{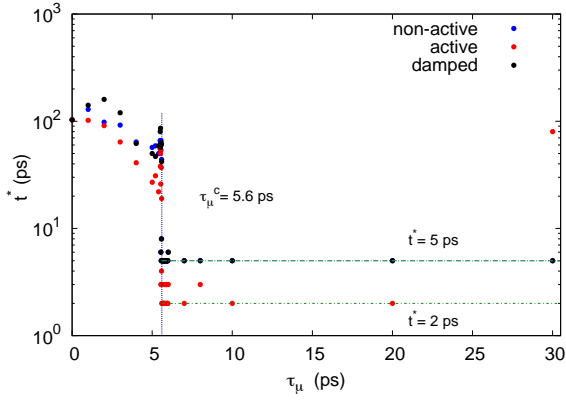}
\caption{{\color{black}(color online)  Non Gaussian parameter $\alpha_{2}(t)$ characteristic time $t^{*}$ versus $\tau_{\mu}$ parameter. } }
\label{falt}
\end{figure}

\begin{figure}%[H]
\centering
\includegraphics[height=6.8 cm]{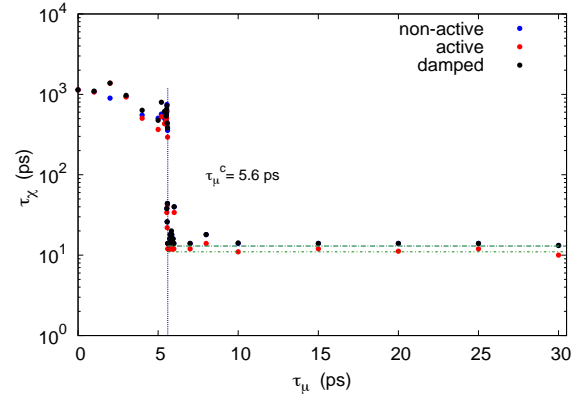}
\caption{{\color{black}(color online)  Characteristic time ${\tau}_{\chi}$ of our dynamic susceptibility, defined as the time for which $\chi_{4}(a_{1},t)$ reaches its maximum value,  versus $\tau_{\mu}$.   $a_{1}=2$\AA\ (the optimum above the transition).} }
\label{fk3b}
\end{figure}

We observe a transition at  $ \tau_{\mu}^{c}=5.6 ps$ for the two characteristic times studied, the $\alpha$ relaxation time $\tau_{\alpha}$ (Figure \ref{ftaualpha}) characterizing the liquid local dynamics, and the characteristic time $t^{*}$ that characterizes the heterogeneous dynamics (Figure \ref{falt}), and the characteristic time $\tau_{\chi}$ of the susceptibility  in Figure \ref{fk3b}.

Above the transition\cite{transition}, we find $\tau_{\alpha}\approx 2 \tau_{\mu}^{c}$ and $t^{*} \approx  \tau_{\mu}^{c}$.
If one expects that $\tau_{\alpha}>t^{*}$ as $\tau_{\alpha}$ corresponds to the complete relaxation of the medium, while $t^{*}$ corresponds to the very beginning of the cage escaping process that leads eventually to the relaxation, the fact that  $t^{*} \approx  \tau_{\mu}^{c}$ is however of particular interest, as it suggests that  $\tau_{\mu}$ interacts with the medium's cooperativity.

\begin{figure}%[H]
\centering
\includegraphics[height=6.8 cm]{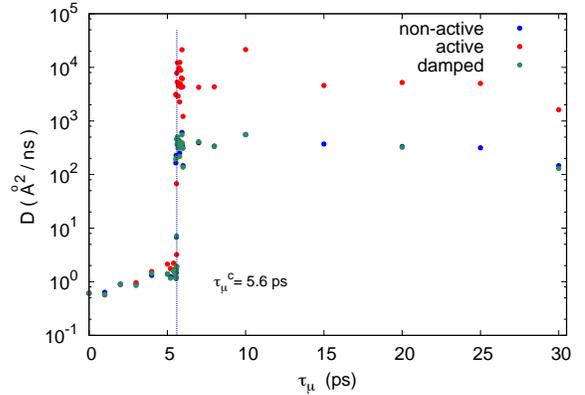}
\caption{{\color{black}(color online)  Diffusion coefficient $D$ versus $\tau_{\mu}$ where $\tau_{\mu}$ is the characteristic time chosen for the mobility definition acting in the activation force. } }
\label{fdiffu}
\end{figure}

The same transition appears for the diffusion coefficient $D$ (Figure \ref{fdiffu}) and as a result to the associated characteristic time $\displaystyle {\tau_{D}={{b^{2}}\over{D}}}$ where we define the characteristic length $b$ as the average distance necessary to get outside the cage created by the surrounding molecules. %, that we evaluate from the radial distribution function around $2.2$\AA\.

\begin{figure}%[H]
\centering
\includegraphics[height=6.8 cm]{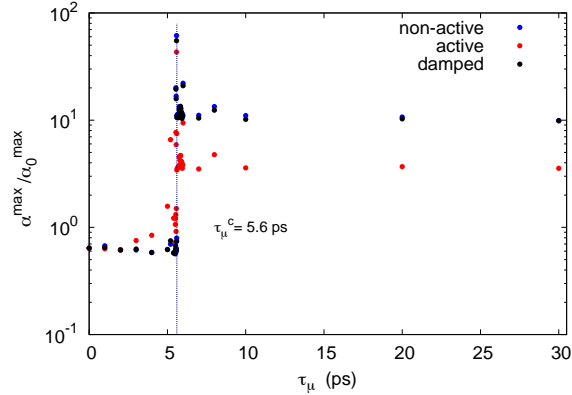}
\caption{{\color{black}(color online)  Non Gaussian parameter $\alpha_{2}(t)$ normalized maximum value versus $\tau_{\mu}$ parameter.
$\tau_{\mu}$ is the characteristic time chosen for the mobility definition acting in the activation force. } }
\label{fal}
\end{figure}

\begin{figure}%[H]
\centering
\includegraphics[height=6.8 cm]{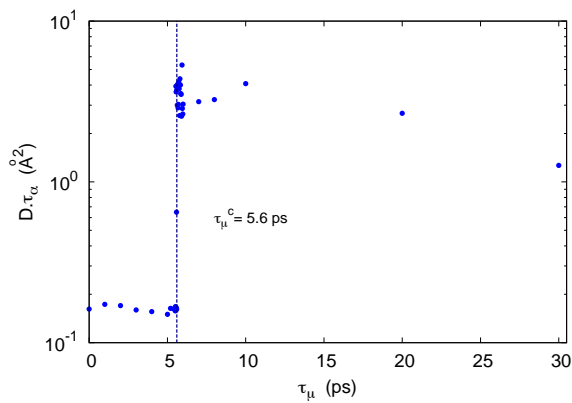}
\caption{{\color{black}(color online)  Breaking of the Stokes-Einstein relation versus $\tau_{\mu}$ parameter. } }
\label{fse}
\end{figure}

\begin{figure}%[H]
\centering
\includegraphics[height=6.8 cm]{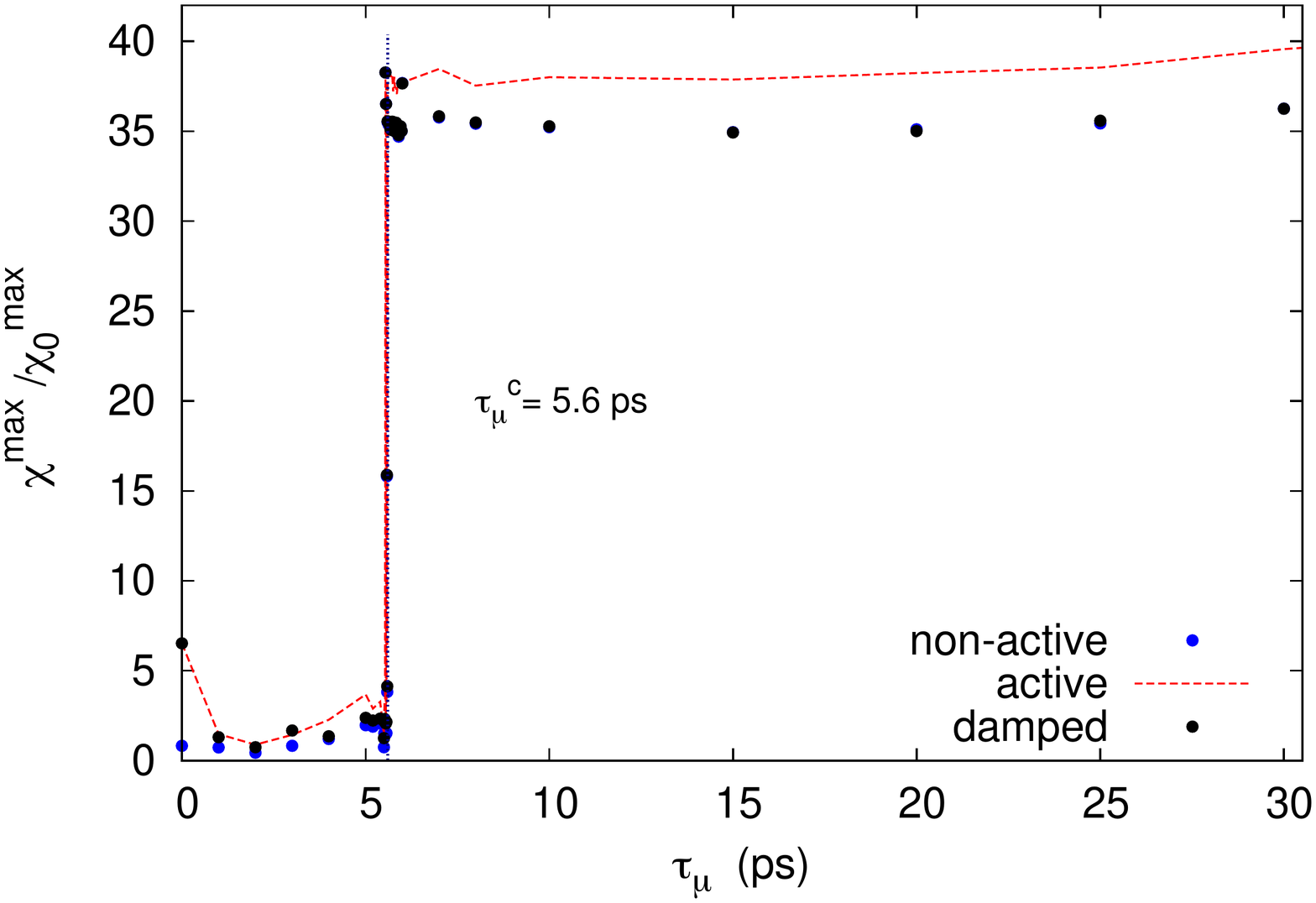}
\caption{{\color{black}(color online)  Dynamic susceptibility  maximum value $\chi_{4}(a_{1},t)^{max}$ normalized by the maximum value in the non-active liquid $\chi_{4}(a_{0},t)_{0}^{max}$ versus $\tau_{\mu}$. For clarity, we have chosen here  $a_{1}=2$\AA\ (the optimum above the transition) and $a_{0}=1$\AA\ (the optimum value without activation and below the transition) for the whole set of points.} }
\label{fk3}
\end{figure}

When $\tau_{\mu}$ increases, the characteristic times first decrease continuously then drop to a constant value as $ \tau_{\mu}^{c}=5.6 ps$ is reached.
Notice that at the transition we observe for the whole set of parameters a peak corresponding to an increase of the characteristic times ($t^{*}$ and $\tau_{\alpha}$), while the diffusion coefficient unexpectedly increases ($\tau_{D}$ decreases).
The origin of this slowing down located at the transition, just before the large fluidization, appears related to a cooperativity increase.  We observe indeed a similar peak for  the Non Gaussian parameter (Figure \ref{fal}) and for the deviation from the Stokes Einstein law (Figure \ref{fse}) showing that the dynamic is highly cooperative at that point, as expected for a phase transition\cite{Chandler,Pathria,Gould}. The dynamic susceptibility for non-active molecules displays also a small peak that reaches the active value in Figure \ref{fk3}.

\subsection{Increase of dynamical heterogeneity upon activation\\}

Dynamical heterogeneity (DH)\cite{dh0,dh1,dh2,dh3}, are together with a dramatic increase of the medium's viscosity, a hallmark of supercooled liquids in their approach to the glass transition. 
They are characterized by the spontaneous aggregation of most mobile molecules on a characteristic time $t^{*}$ and string like cooperative motions of these molecules, on the same characteristic time.
As cooperative mechanisms are expected with a rising associated susceptibility in any phase transition, these cooperative mechanisms have long been suspected to be the fingerprint of a thermodynamic phase transition explaining the glass transition.  Also, the DHs are a crucial element in facilitation theories\cite{facile,facile1,facile2,facile3,facile4,facile5,facile6}.  

\subsubsection{Dynamic heterogeneity\\}

 %The dynamic susceptibility $\chi_{4}(t)$ is one of the most efficient parameter devised to measure the dynamic heterogeneity in supercooled liquids, and we will therefore first use $\chi_{4}(t)$ to study the behavior of the DHs upon activation of our medium, then we will study the behavior of other significant parameters as the Non Gaussian parameter and the breaking of the Stokes-Einstein law.

\begin{figure}%[H]
\centering
\includegraphics[height=6.2 cm]{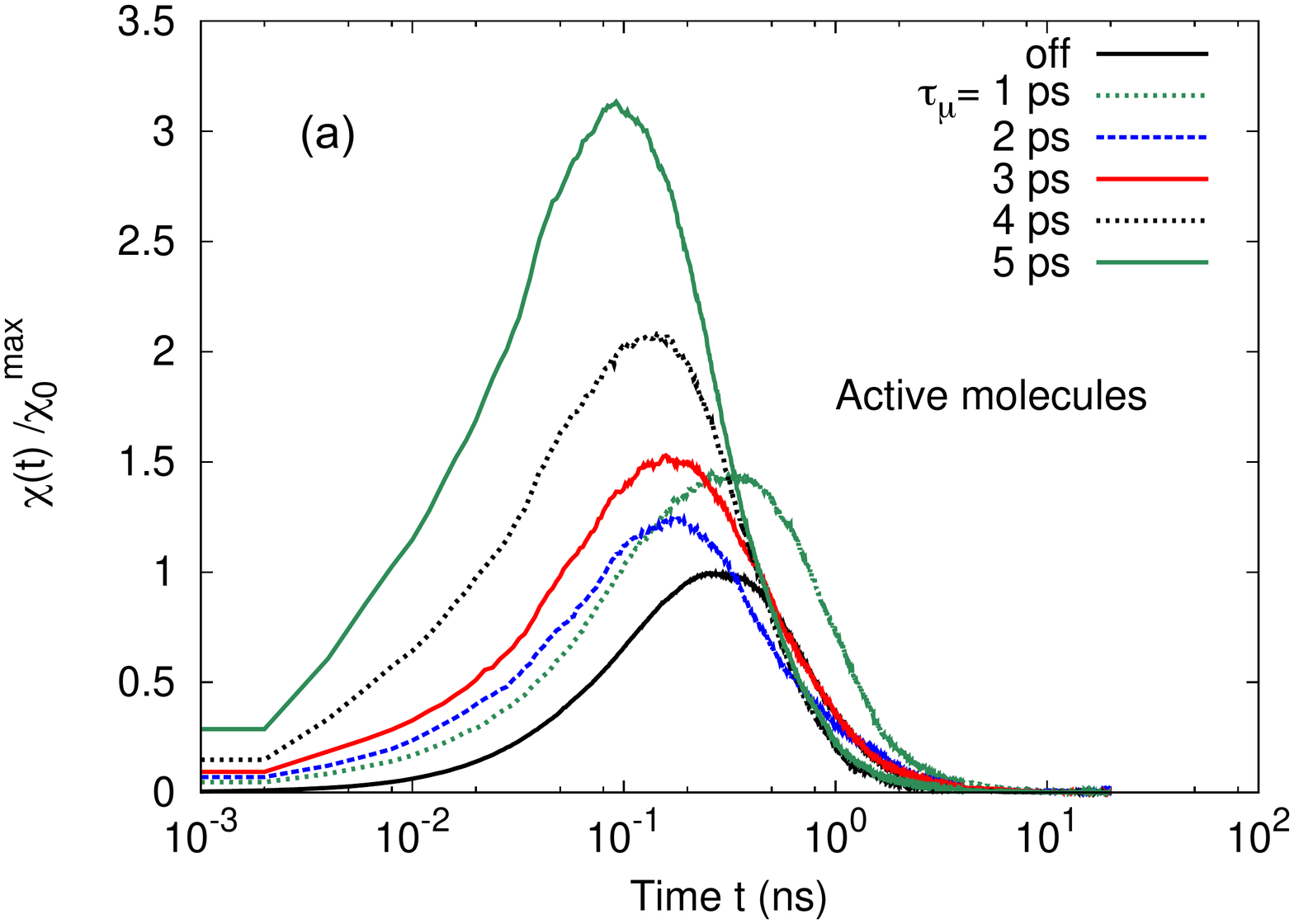}
\includegraphics[height=6.2 cm]{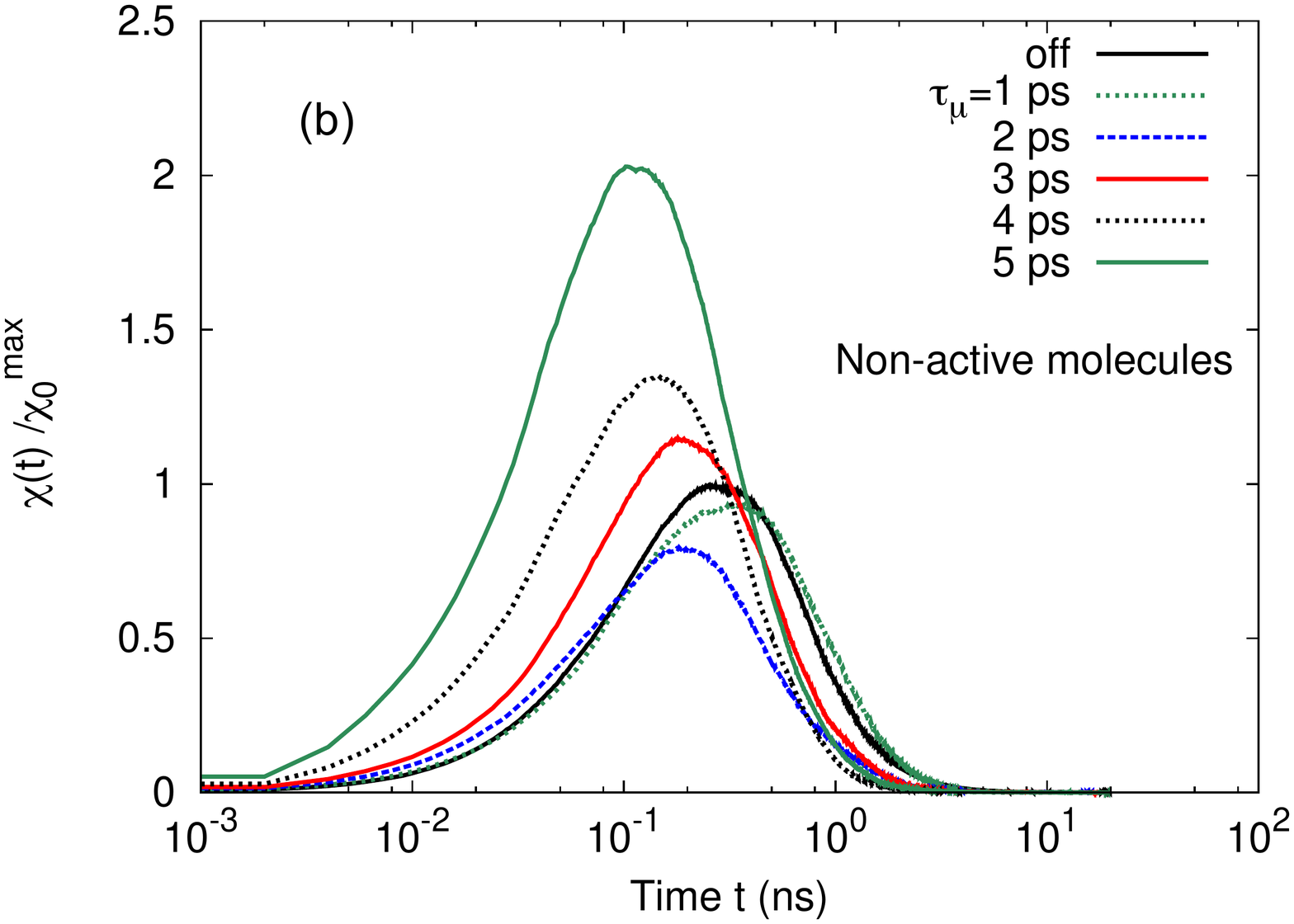}
\includegraphics[height=6.2 cm]{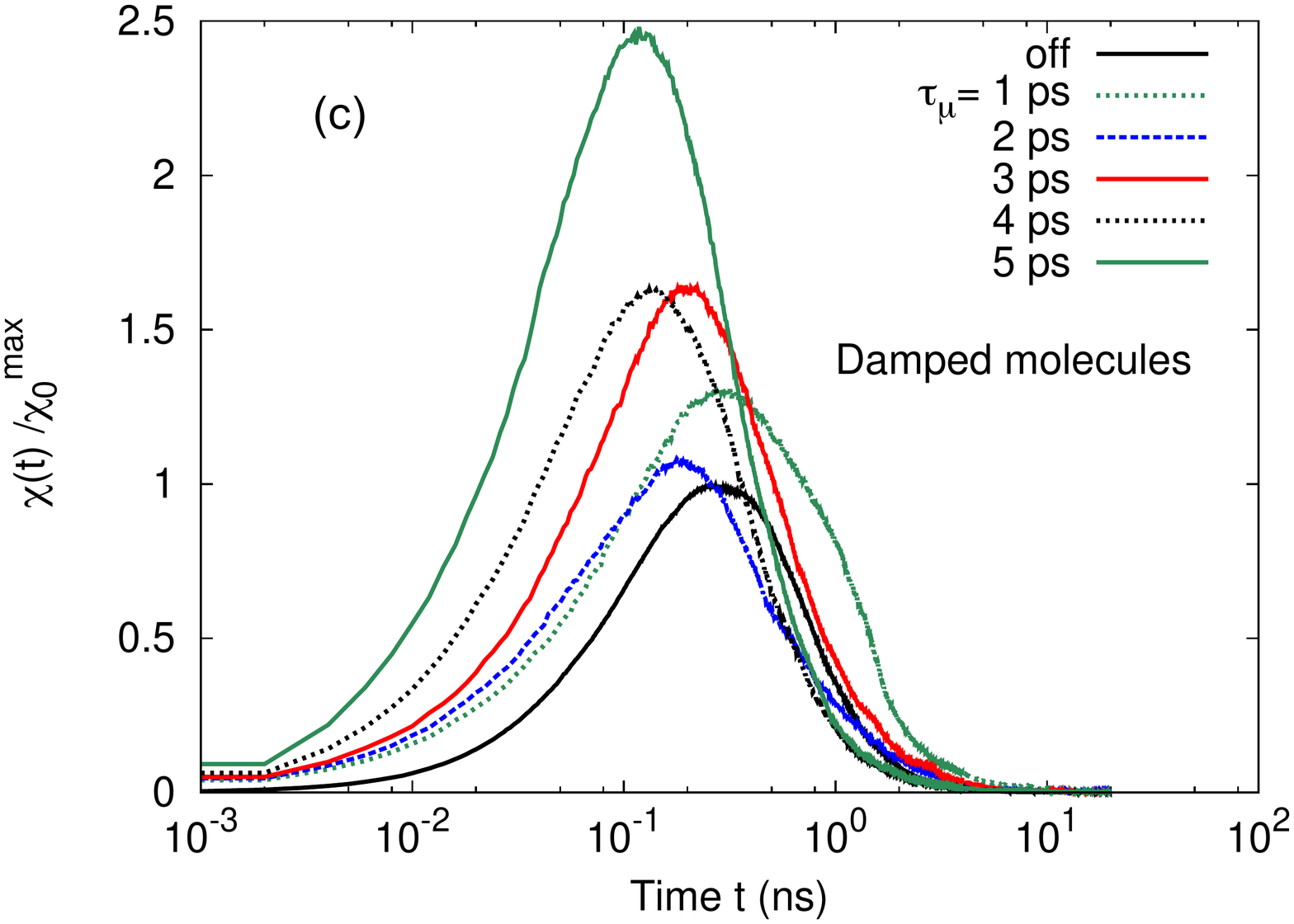}

\caption{{\color{black}(color online)  Dynamic susceptibility $\chi_{4}(a_{0},t)$ normalized by its maximum value in the non-active liquid  $\chi_{4}(a_{0},t)_{0}^{max}$ versus $\tau_{\mu}$ parameter. $\chi_{4}(a_{0},t)/\chi_{4}(a_{0},t)_{0}^{max}$ is shown for (a) the $10$ percent intermittently  active molecules,  (b) the $80$ percent medium molecules moving freely, (c) the $10$ percent intermittently damped molecules.  $a_{0}=1$\AA\ which correspond to the optimum below the transition and for the non-activated liquid.
$\tau_{\mu}$ is the characteristic time chosen for the mobility definition acting in the activation force. Note that in the vicinity of the transition the susceptibility is much larger.} }

\label{fk1}
\end{figure}

\begin{figure}%[H]
\centering
\includegraphics[height=6.2 cm]{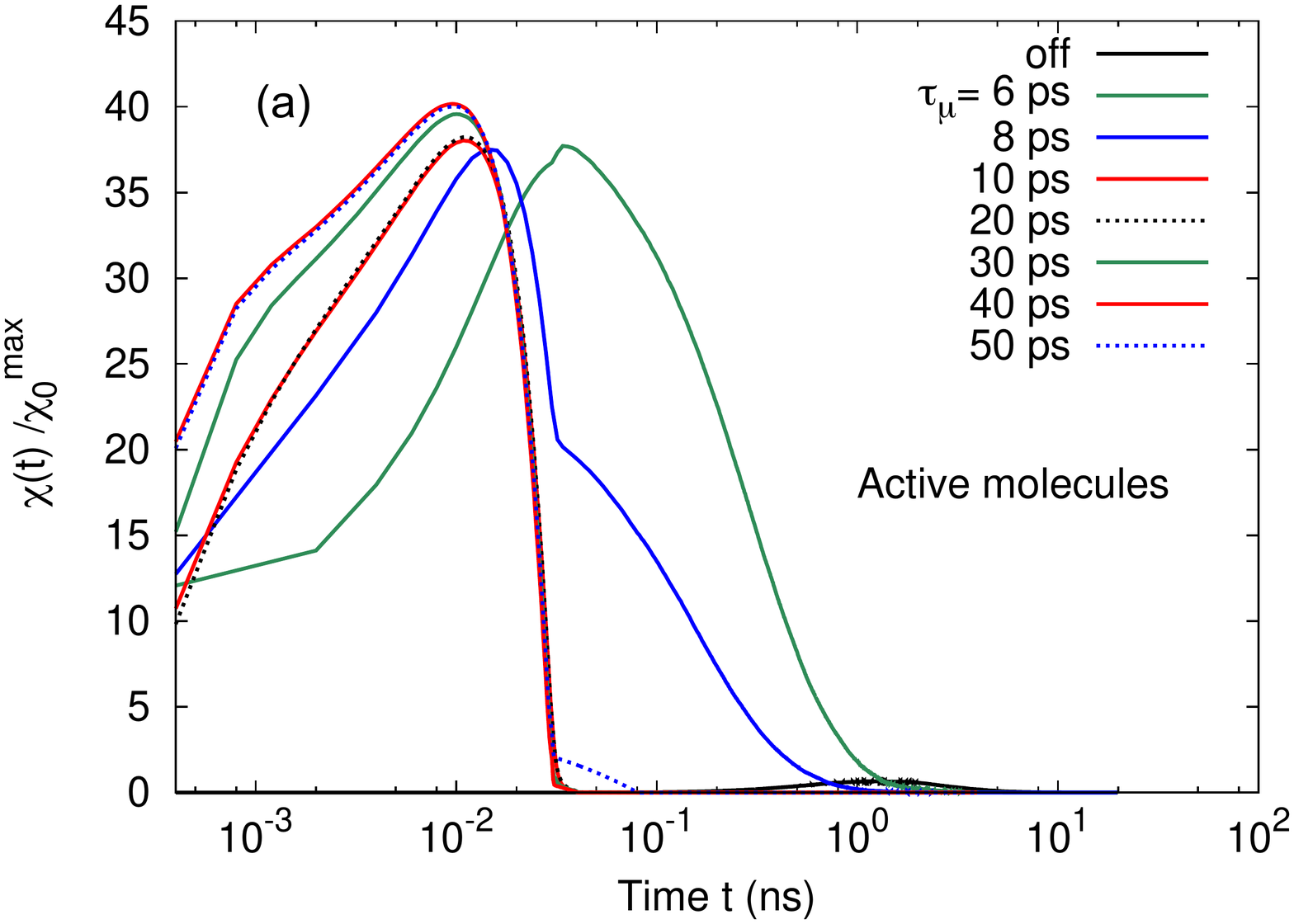}
\includegraphics[height=6.2 cm]{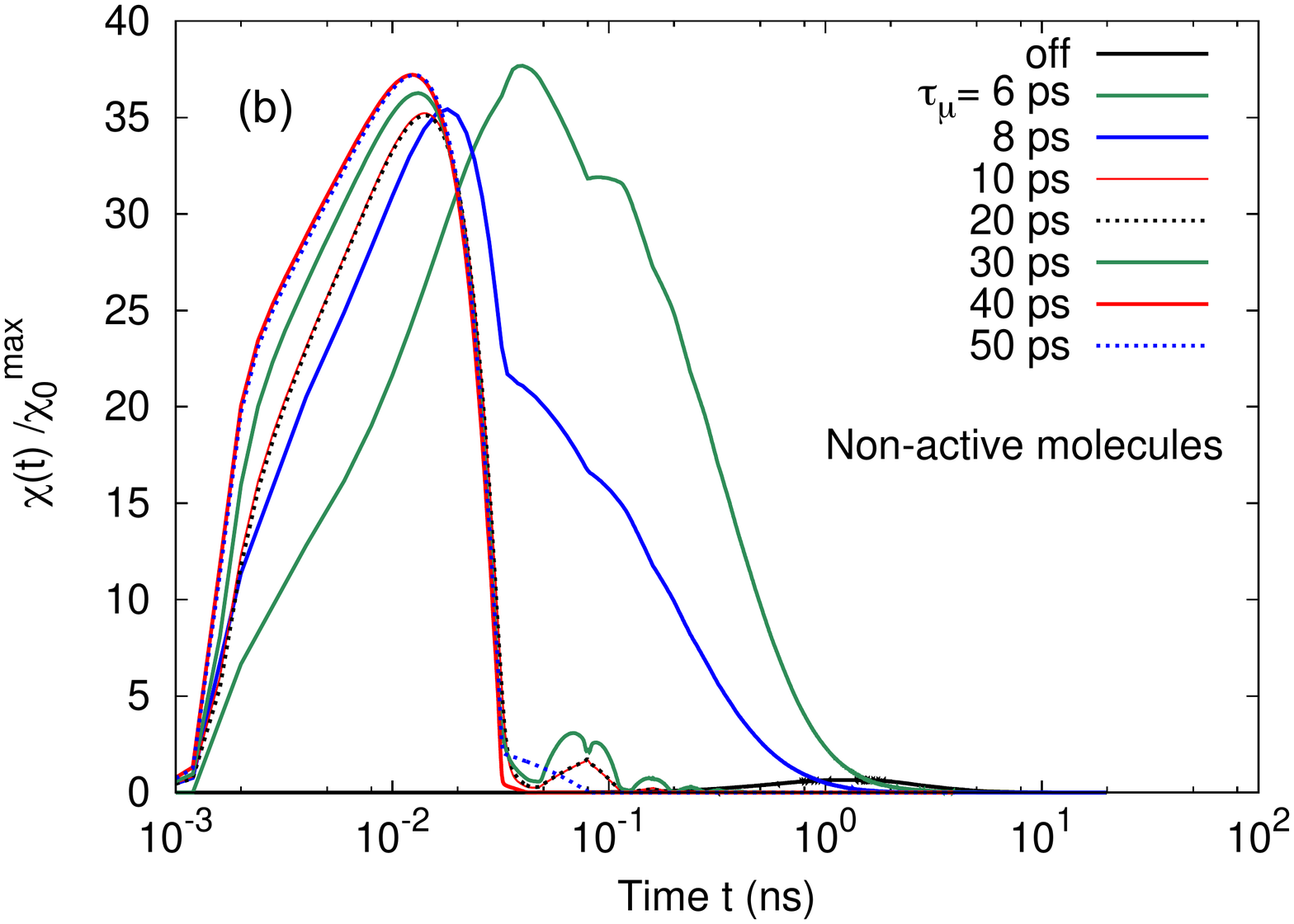}
\includegraphics[height=6.2 cm]{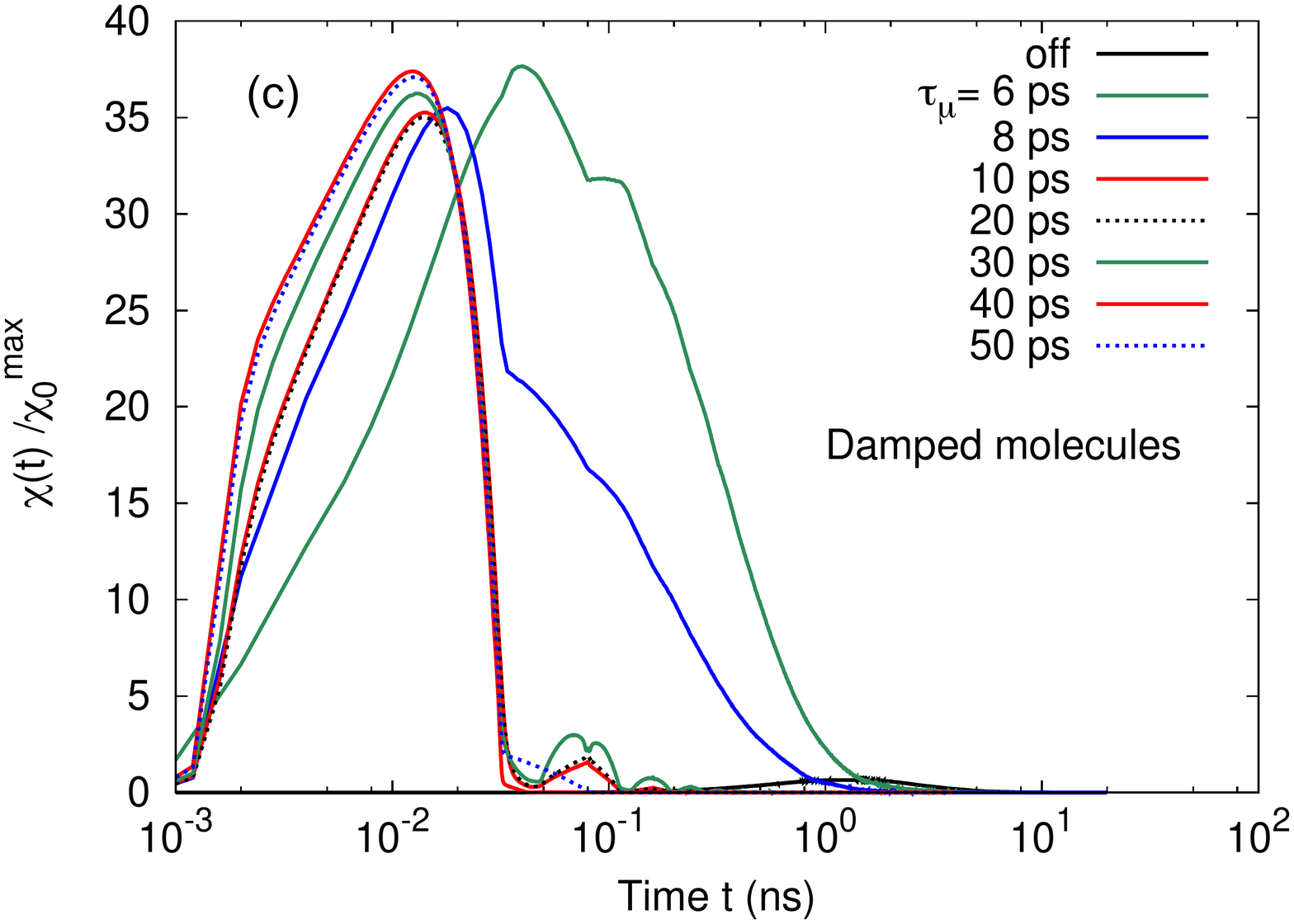}

\caption{{\color{black}(color online)  Dynamic susceptibility $\chi_{4}(a_{1},t)$ normalized by its maximum value in the non-active liquid  $\chi_{4}(a_{0},t)^{max}_{0}$ versus $\tau_{\mu}$ parameter. 
$\chi_{4}(a_{1},t)/\chi_{4}(a_{0},t)_{0}^{max}$ is shown for (a) the $10$ percent intermittently  active molecules, (b) the $80$ percent medium molecules moving freely, (c) the $10$ percent intermittently damped molecules.
$a_{1}=2$\AA\ which correspond to the optimum above the transition, and $a_{0}=1$\AA\ which correspond to the optimum below the transition and for the non-activated liquid.
$\tau_{\mu}$ is the characteristic time chosen for the mobility definition acting in the activation force. Note that in the vicinity of the transition the susceptibility is slightly different and shifted in time.} }
\label{fk2}
\end{figure}

%{\color{black}
Below the phase transition\cite{transition} i.e. for $\tau_{\mu}<{\tau_{\mu}}^{c} $ after a small decrease of the dynamic susceptibility for $\tau_{\mu}=1 $ to $ 2 ps$ we observe in Figure \ref{fk1} a significant increase of the susceptibility with $\tau_{\mu}$ leading to a value equal to twice the thermal susceptibility $\chi_{4}(a_{0},t)^{max}_{0}$ for $\tau_{\mu}={\tau_{\mu}}^{c} $. We also observe a small decrease of the susceptibility characteristic time.
The maximum value of the susceptibility is shifted to shorter times, i.e. $\tau_{\chi}$ that maximizes  $\chi_{4}(a,t)$, decreases. 
 Figure \ref{fk1} shows that $\tau_{\chi}$ decreases from $250 ps$ without activation to $100ps$ for $\tau_{\mu}=5 ps$ just before the transition.
Above the transition i.e. for $\tau_{\mu}>{\tau_{\mu}}^{c} $ (Figure \ref{fk2}), the susceptibility behavior changes drastically and we observe an huge increase of the susceptibility, leading to a value $25$ times larger than the thermal susceptibility. 
The susceptibility characteristic time also undergoes a strong evolution above the transition, decreasing from $250ps$ for the thermal susceptibility to values around $10ps$ .

Figure \ref{fk3} resume that behavior showing an abrupt transition on the susceptibility for $\tau_{\mu} \approx {\tau_{\mu}}^{c}  $.
This Figure compares well with Figures \ref{fdiffu} and \ref{ftaualpha} displaying a similar abrupt transition for the diffusion coefficient and the alpha relaxation time for the same value of $\tau_{\mu} $. 

{\color{black}We therefore observe a phase transition controlled by the mobility time parameter  $\tau_{\mu}$  showing  a large increase of the cooperative behavior (DH) associated to the fluidization of the medium.}
The association of an increase of dynamic heterogeneity with a decrease of the viscosity while quite unusual has been observed in fluidization processes by activation of materials with molecular motors. 

{\color{black} As a tentative picture, in these activated systems the activation induces the dynamic heterogeneity, cooperative motions that in turn induce the fluidization (i.e. a decrease of the viscosity and an increase of the diffusion coefficient). }

%While in non activated supercooled liquids, spontaneous dynamic heterogeneity arise (Le Chatelier principle) as a resistance  conter-reaction to the increase of the viscosity when the temperature drops.

What is the difference in the cooperative behavior of active, non-active and damped molecules subsets ?
Figures \ref{fk1} shows that below the transition, the maximum difference is observed for $\tau_{\mu}=5 ps$ in the vicinity of the transition.
We find the normalized susceptibility to be larger for active molecules than damped molecules and larger for damped molecules than non-active ones.
However the difference is relatively small.
While above the transition the difference is even smaller in relative values as shown in Figure \ref{fk2}.
Therefore, the differences observed for active, non-active and damped molecules for the susceptibility are small, showing that the whole medium's cooperativity is affected by active molecules motion.

%\vskip0.5cm
The large increase of the DHs at the transition observed with the dynamic susceptibility are confirmed with other measures of the dynamic heterogeneity, as the Non Gaussian parameter (Figures \ref{fal} and \ref{falt}) and the breaking of the Stokes-Einstein law (Figure \ref{fse}).
%For active molecules below the phase transition we observe for the NGP a behavior similar to what we observe for the susceptibility.
%The NGP  first decreases slightly then increases rapidly to reach $2.5$ times its thermal value for $\tau_{\mu}=5ps$ (in qualitative agreement with the factor 2 obtained for the whole set of molecules susceptibility).
%Above the transition as for the susceptibility $\chi(t)$ we find a strong increase of $\alpha_{2}(t^{*})$ but smaller than for $\chi(t_{m})$ (we reach a factor 14 instead of 25).
%We also find a drop in the characteristic time $t^{*}$ as for the susceptibility for $t_{\chi}$.
%Therefore the NGP for active molecules follows qualitatively the same behavior than the global susceptibility.
%Lets turn now to the damped and inactive molecules that will appear to have very similar behavior. Below the transition the inactive and damped molecules NGP stays around (mostly slightly below) its thermal value, showing only slight effects on the DHs.
%Above the transition however both NGP increase hugely to a factor $33$ at the transition and then decreases for larger $\tau_{\mu}$.
%Meanwhile, the characteristic time $t^{*}$ drops to $8-10ps$.
%Interestingly enough this NGP is maximum when $\tau_{\mu} \approx t^{*}$ then decreases for larger values  ($\tau_{\mu}>t^{*}$).
To summarize, we observe for all the statistical functions considered ($\alpha_{2}(t)$, $\chi_{4}(t)$, $D.\tau_{\alpha}$, $D$ and $\tau_{\alpha}$) an important reaction of the system when the mobility used for the activation of molecules corresponds to the physical mobility of the medium.

{\color{black}
\section{Interpretation}

One expects active molecules to induce a larger mobility around them while damped molecules hinder the motions in their vicinity.
That effect is amplified in supercooled liquids due to the rise of cooperativity.
However in our system some additional phenomena take place.
If an active molecule pass in the vicinity of another one, their mobilities will tend to align due to the definition of our force.
After this chance encounter the active molecules will therefore have a larger probability to stay in vicinity due to relatively similar mobilities.
This leads ultimately to an aggregation of active molecules and a phase transition with fluidization of the medium.
However for the transition to take place the force has to be in the direction of a physically relevant mobility leading to a threshold  for $\tau_{\mu} \approx t^{*}$.
}
\section{Conclusion}

An Induced fluidization was  reported\cite{flu1,flu2,flu3,flu4,flu5,md16,cage} experimentally and with simulation when simple molecular motors\cite{az1,az2,az3,az4,az5,az6,az7,az8,az9,az10} activate a soft material.
That induced fluidization was interpreted to be due to the activation of the spontaneous (thermal) cooperative mechanisms in supercooled mediums.
Molecules pushed away by the nano-motor induce molecular motions around them due to the medium's cooperativity. 

Here we raised the question of inducing new cooperative mechanisms inside the medium using active molecules with an activation that depended on the mobility of their surroundings.
We expected the facilitation created in that way to induce cooperativity resulting in a fluidization of the medium.
We found actually that the cooperativity (Dynamic heterogeneity) of the medium increased hugely upon activation.
Together with that induced cooperativity we found an important fluidization of the medium.

{\color{black}
The created facilitation mechanism followed by the active particles was found to induce a dynamic phase transition in the whole liquid when the characteristic time used for the mobility definition reaches a critical value $\tau_{\mu}^{c}$. That transition is characterized by an aggregation of the active molecules, a large increase of the dynamic susceptibility and a drop of the characteristic times and relaxation times that is a fluidization of the medium.
After the fluidization takes place, the DH characteristic time $t^{*}$ equals the critical value $\tau_{\mu}^{c}$ suggesting an interplay between the medium's spontaneous cooperativity and the active molecules artificial facilitation. 

At the transition we observe a peak of the susceptibility, Non-Gaussian parameter, Stokes-Einstein's law deviation, relaxation time $\tau_{\alpha}$,  DH characteristic time $t^{*}$ and diffusion coefficient.
The increase of the viscosity implied by the relaxation times increase on the peak seems at first sight in contradiction with the increase of the diffusion coefficient.
However the large increase of the cooperativity and the aggregation of mobile molecules that favor diffusion upon viscosity could explain that behavior.
}

\vskip 0.5cm

{\bf \Large  Conflict of interest}

There are no conflict of interest to declare.

\vskip 0.5cm

%{\bf \Large  Data availability}

%The data that support the findings of that study are available from the corresponding author upon reasonable request.

%\vskip 0.5cm


\begin{thebibliography}{99}
\bibitem{transition} In this paper we call respectively "above or below the transition" the medium that has undergone the phase transition or not.
The reason is that our main parameter $\tau_{\mu}$ is larger after the transition than before. Thus "above" or "below" refer to that critical parameter $\tau_{\mu}={\tau_{\mu}}^{c}  $.
\end{thebibliography}
\end{document}